\documentclass[a4paper]{article}
\usepackage{graphicx}
\newcommand{\UA}{\mbox{$U_{A}(1)$}}
\newcommand{\qbar}{\mbox{$\bar q$}}
\newcommand{\ubar}{\mbox{$\bar u$}}
\newcommand{\dbar}{\mbox{$\bar d$}}
\newcommand{\sbar}{\mbox{$\bar s$}}
\newcommand{\ie}{{\it i.e.}}
\newcommand{\bea}{\begin{eqnarray}}
\newcommand{\eea}{\end{eqnarray}}
\newcommand{\la}{\left\langle}
\newcommand{\ra}{\right\rangle}

\begin{document}
\begin{center}
{\Large \bf $U_A(1)$ Breaking Effects on the Light Scalar Meson Spectrum}
\vskip 10mm
Kenichi \textsc{Naito}$^{1,}$
\footnote{E-mail: knaito@nucl.sci.hokudai.ac.jp}, 
Makoto \textsc{Oka}$^{2,}$\footnote{E-mail: oka@th.phys.titech.ac.jp},
Makoto \textsc{Takizawa}$^{3,}$
\footnote{E-mail: takizawa@ac.shoyaku.ac.jp} \\ 
and Toru \textsc{Umekawa}$^{2,}$
\footnote{E-mail: umekawa@th.phys.titech.ac.jp}
\vskip 2mm
$^{1}$Meme Media Laboratory, Graduate school of Enginieering, 
Hokkaido University,\\ Sapporo, Hokkaido 060-8628 Japan\\
$^{2}$Department of Physics, Tokyo Institute of Technology, Meguro, 
Tokyo 152-8551 Japan\\
$^{3}$Showa Pharmaceutical University, Machida, Tokyo 194-8543 Japan
\end{center}
\vskip 10mm
\begin{abstract}
Effects of the \UA\ breaking interaction on the low-lying 
nonet scalar mesons are studied using the extended Nambu-Jona-Lasinio model. 
The strength of the \UA\ breaking interaction is determined by the 
electromagnetic decays of the $\eta$ meson. We find that the 
\UA\ breaking interaction gives rise to about 150 MeV
mass difference between the $\sigma$ and $a_0$ mesons. 
We also find that the strangeness content in the $\sigma$ meson is 
about 15\%. 
The calculated mass of the $I = 1/2$ state is about 200 MeV heavier than 
that of the $I = 1$ state.  The order of these masses is not likely to change
within this model.
\end{abstract}
\section{Introduction}
Understanding low-lying hadron spectrum is one of the most challenging 
problems in the quantum chromodynamics (QCD).  
The spectrum is highly nontrivial due to the nonperturbative 
complexity, such as spontaneous breaking of chiral symmetry and
axial U(1) anomaly.  
For instance, the pseudoscalar mesons are off-scale light, if we 
suppose that they are bound states of a quark and an antiquark, 
while their spin-flip partners, \ie, vector mesons, are normal with
masses about 2/3 of the baryon masses.
This ``anomaly'' in the pseudoscalar mesons is attributed to 
their Nambu-Goldstone-boson nature associated with the 
spontaneous chiral symmetry breaking.
This is strongly in contrast with the heavy meson spectroscopy,
such as heavy quarkonia and heavy flavor mesons.  There the 
spectrum is more like the hydrogen atom with slightly
stronger fine and hyperfine splittings.

It should be noticed that the low-lying hadrons are the key to
explore the complicated QCD vacuum, as in QCD we are not able to
``measure'' the bulk properties of the ground state, which 
can be accessed directly in the case of condensed matter physics.
Thus it is important to explore the properties of the low-lying
hadrons from the viewpoints of QCD dynamics and symmetries.

Another nontrivial effect comes from \UA\ symmetry, which is
expected to be broken by anomaly.
Weinberg showed that the mass of $\eta'$ should be less
than $\sqrt{3} m_{\pi}$ if \UA\ symmetry were not
explicitly broken \cite{Weinberg1975}.
Thus the \UA\ symmetry must be broken.
In the following year, 't Hooft pointed out the relation between 
\UA\ anomaly and topological gluon configurations of QCD and 
showed that the interaction of light quarks and instantons breaks 
the \UA\ symmetry \cite{tHooft1976}.
He also showed that such an interaction can be represented by a
local $2N_{f}$ quark vertex, which is antisymmetric under
flavor exchanges, in the dilute instanton gas approximation.
The dynamics of instantons in the multi-instanton vacuum has
been studied by many authors, either in the models or in the
lattice QCD approach, and the widely accepted picture is that
the QCD vacuum consists of small instantons of the size about
1/3 fm with the density of 1 instanton (or anti-instanton) per
fm$^{4}$ \cite{Instanton}.

According to such an instanton vacuum picture, the hadron spectrum 
shows its signature.  The $\eta-\eta'$ mass difference is the obvious one, 
which can be understood by flavor mixing in the $I=0$ 
$(q\qbar)_{NS} \equiv {1\over \sqrt{2}} \left( u\ubar+d\dbar \right)$ 
and $s\sbar$. 
Without the flavor mixing, $(q\qbar)_{NS}$ and $s\sbar$ would form 
mass eigenstates, and thus the ideal mixing is achieved. 
This is natural if the Okubo-Zweig-Iizuka (OZI) rule applies.
However, the OZI rule is known to be significantly broken in 
the pseudoscalar mesons.
For instance, according to our previous analyses in the 
Nambu-Jona-Lasinio model,  the electromagnetic $\eta$ decay 
processes  indicate that the mixing of $(q\qbar)_{NS}$ 
and $s\sbar$ is indeed strong so that the $\eta$ meson is close to  
the pure octet state \cite{TNO1997}.

Recently, the scalar mesons, $J^{\pi}=0^{+}$, attract a lot of 
attention by two reasons \cite{sigmaworkshop}.
(1) Experimental evidence for $\sigma$ 
($I=0$) scalar meson of mass around 500-800 MeV is overwhelming \cite{PDG2002}.
Especially the decays of heavy mesons show clear peaks in the 
$\pi\pi$ invariant mass spectrum.
(2) The roles of the scalar mesons in chiral symmetry have been 
stressed in the context of high temperature and/or density hadronic 
matter \cite{HK1985}.
It is believed that chiral symmetry will be restored in the QCD ground 
state at high temperature (and/or baryon density).  Above the critical 
temperature of order 150 MeV the world is nearly chiral symmetric and 
we expect that hadrons belong to irreducible representations of 
chiral symmetry, if we neglect small mixing due to finite quark mass.
The pion is not any more a Nambu-Goldstone boson, and has a finite 
mass and should be degenerate with a scalar meson, \ie, sigma.

In this paper, we study the masses and mixing angles of scalar 
mesons in the context of chiral symmetry and \UA\ breaking using 
the extended NJL model, in which the SU(3) NJL model is supplemented 
with the Kobayashi-Maskawa-'t Hooft (KMT) determinant interaction 
\cite{tHooft1976,KM1970}.
This is the simplest possible quark model with the correct symmetry 
structure for the present purposes.  The chiral symmetry is broken 
both explicitly by a quark mass term and dynamically by quark loops,
while \UA\ symmetry is broken by the KMT interaction term.

Why is the \UA\ expected to be important in the scalar mesons?  
It is because the KMT interaction selects out the scalar sector as 
well as the pseudoscalar mesons and therefore the OZI rule may be 
significantly broken also in the scalar mesons.

Recent experimental data suggest that the light scalar mesons (below 1GeV)
show strange mass patterns, \ie, 
$\sigma (600) - \kappa (700-900) - f_{0}(980)-a_{0}(980)$.
\footnote{The existence of a light and very broad $\kappa$ meson is 
controversial. It is not our aim to claim the existence of the 
light $\kappa$ meson.}
This pattern cannot be explained by a  $q\qbar$ nonet, 
because the $I=1$ $a_{0}$ states are degenerate with the 
second $I=0$ state $f_{0}$, while the first $I=0$ $\sigma$ is far below them.
Furthermore the strange meson $\kappa$ comes below $a_0$.

Dmitrasinovic \cite{Dm1996} has pointed out that the KMT interaction may 
change the spectrum of the scalar mesons significantly.  He has shown that 
the mass splitting of the $I=0$ and $I=1$ nonstrange scalar mesons is 
generated by the KMT interaction.  He, however, tried to assign the 
$I=0$ state to $f_{0}(980)$ or higher, and the $I=1$ to $a_0(1450)$.
Thus, his results are not applied to the lower mass mesons phenomenologically.

We construct the model with the \UA\ anomaly which is strong enough to explain
the $\eta$ decay widths and apply it to the light scalar mesons.  
We will obtain significant $\sigma-a_{0}$ splitting and flavor 
mixing from the KMT interaction.
The mixing of $s\sbar$ component in the $\sigma$ meson is very interesting.
As the extended NJL model has been used in the analyses of the pseudoscalar 
mesons, it has an advantage that the parameters have been all fixed in 
the pseudoscalar sector.  

In section 2, we present the extended NJL model.  
The formulation of solving the Bethe-Salpeter equation 
for the scalar channel is explained.
In section 3, we show our results and give discussions on the mass 
spectrum as well as the mixings.
In section 4, conclusion and future prospects are given.
\section{Formulation}
We work with the following NJL model lagrangian density extended to 
three-flavor case:
\bea
{\cal L} & = & {\cal L}_0 + {\cal L}_4 + {\cal L}_6 , \label{njl1} \\
{\cal L}_0 & = & \bar \psi \,\left( i \partial_\mu \gamma^\mu - \hat m 
\right) \, \psi \, ,
\label{njl2} \\
{\cal L}_4 & = & {G_S \over 2} \sum_{a=0}^8 \, \left[\, \left( 
\bar \psi \lambda^a \psi \right)^2 + \left( \bar \psi \lambda^a i \gamma_5 
\psi \right)^2 \, \right] \, ,
\label{njl3} \\
{\cal L}_6 & = & G_D \left\{ \, {\rm det} \left[ \bar \psi_i (1 - \gamma_5) 
\psi_j \right] + {\rm det}  \left[ \bar \psi_i (1 + \gamma_5) \psi_j 
\right] \, \right\} \, .
\label{njl4}
\eea
Here the quark field $\psi$ is a column vector in color, flavor and Dirac 
spaces and $\lambda^a (a=0\ldots 8)$ is the Gell-Mann matrices for 
the flavor $U(3)$. 
The free Dirac lagrangian ${\cal L}_0$ incorporates the current quark mass 
matrix $\hat m = {\rm diag}(m_u, m_d, m_s)$ which breaks the chiral 
$U_L(3) \times U_R(3)$ invariance explicitly. ${\cal L}_4$ is a QCD 
motivated four-fermion interaction, which is chiral $U_L(3) \times U_R(3)$
invariant.  The Kobayashi-Maskawa-'t Hooft determinant ${\cal L}_6$ 
represents the $U_A(1)$ anomaly.  
It is a $3 \times 3$ determinant with respect to flavor with 
$i,j = {\rm u,d,s}$.   

Quark condensates and constituent quark masses are self-consistently 
determinded by the gap equations in the mean field approximation,
\bea
M_u & = & m_u - 2G_S \la \ubar u \ra - 
  2 G_D \la \dbar d \ra \la \sbar s \ra \, , \nonumber \\
M_d & = & m_d - 2G_S \la \dbar d \ra - 
  2 G_D \la \sbar s \ra \la \ubar u \ra \, , \nonumber \\
M_s & = & m_s - 2G_S \la \sbar s \ra - 
  2 G_D \la \ubar u \ra \la \dbar d \ra \, , \label{gap}
\eea
with 
\bea
\la \qbar q \ra & = & - {\rm Tr}^{(c,D)} \left[ iS_F^q (x = 0) \right] 
\nonumber \\
& = & - \int^\Lambda \frac{d^4p}{(2\pi)^4} {\rm Tr}^{(c,D)}
\left[ \frac{i}{p_\mu \gamma^\mu - M_q + i\epsilon} \right] \, .
\label{condensate}
\eea
Here the covariant cutoff $\Lambda$ is introduced to regularize the
divergent integral and Tr$^{(c,D)}$ means trace in color and Dirac spaces.

The scalar channel quark-antiquark scattering amplitudes
\begin{equation}
\la p_3 , \bar p_4 ; {\rm out} \right. \left| p_1 , \bar p_2 ; {\rm in} \ra 
 =  (2 \pi)^4 \delta^4(p_3 + p_4 - p_1 - p_2) {\cal T}_{q \bar q} 
\end{equation}
are then calculated in the ladder approximation. 
We assume that $m_u = m_d$ so that the isospin is exact.
In the $\sigma$ and $f_0$ channel, 
the explicit expression is 
\begin{equation}
{\cal T}_{q \bar q} = -
\left(
\begin{array}{c} 
\bar u(p_3) \lambda^8 v(p_4) \\
\bar u(p_3) \lambda^0 v(p_4) 
\end{array} 
\right)^T \,
\left(
\begin{array}{cc}
A(q^2) & B(q^2) \\
B(q^2) & C(q^2) \\
\end{array}
\right) \,
\left(
\begin{array}{c}
\bar v(p_2) \lambda^8 u(p_1) \\
\bar v(p_2) \lambda^0 u(p_1)
\end{array}
\right) \, , \label{qas1}
\end{equation}
with 
\begin{eqnarray}
A(q^2) & = & \frac{2}{{\rm det}{\bf D}(q^2)} 
\left\{ 2 ( G_0 G_8 - G_m G_m ) I^0 (q^2) - G_8 \right\} \, , \label{qas2} \\
B(q^2) & = & \frac{2}{{\rm det}{\bf D}(q^2)}
\left\{- 2 ( G_0 G_8 - G_m G_m ) I^m (q^2) - G_m \right\} \, , \label{qas3} \\
C(q^2) & = & \frac{2}{{\rm det}{\bf D}(q^2)}
\left\{ 2 ( G_0 G_8 - G_m G_m ) I^8 (q^2) - G_0 \right\} \, , \label{qas4} 
\end{eqnarray}
and 
\bea
G_0 & = & \frac{1}{2} G_S - \frac{1}{3} ( 2 \langle \bar uu \rangle + 
\langle \bar ss \rangle ) G_D \, , \\
G_8 & = & \frac{1}{2} G_S - \frac{1}{6} ( \langle \bar ss \rangle - 4  
\langle \bar uu \rangle ) G_D \, , \\
G_m & = & - \frac{1}{3 \sqrt{2}} ( \langle \bar ss \rangle - 
\langle \bar uu \rangle ) G_D \, .
\eea
The quark-antiquark bubble integrals are defined by
\begin{eqnarray}
I^0(q^2) & = & i \int^{\Lambda} \frac{d^4p}{(2 \pi)^4} {\rm Tr}^{(c,f,D)}
\left[ S_F(p) \lambda^0 S_F(p+q) \lambda^0 \right]
\, , \label{int1} \\
I^8(q^2) & = & i \int^{\Lambda} \frac{d^4p}{(2 \pi)^4} {\rm Tr}^{(c,f,D)}
\left[ S_F(p) \lambda^8 S_F(p+q) \lambda^8 \right]
\, , \label{int2} \\
I^m(q^2) & = & i \int^{\Lambda} \frac{d^4p}{(2 \pi)^4} {\rm Tr}^{(c,f,D)}
\left[ S_F(p) \lambda^0 S_F(p+q) \lambda^8 \right]
\, , \label{int3} 
\end{eqnarray}
with $q = p_1 + p_2$.  The $2 \times 2$ matrix ${\bf D}$ is given by
\begin{equation}
{\bf D}(q^2) = 
\left( 
\begin{array}{cc}
D_{11}(q^2) & D_{12}(q^2) \\
D_{21}(q^2) & D_{22}(q^2) 
\end{array}
\right) \, , \label{mat}
\end{equation}
with
\begin{eqnarray}
D_{11}(q^2) & = & 2 G_0 I^0(q^2) + 2 G_m I^m(q^2) - 1 \, , \label{mat11}\\
D_{12}(q^2) & = & 2 G_0 I^m(q^2) + 2 G_m I^8(q^2) \label{mat12} \\
D_{21}(q^2) & = & 2 G_8 I^m(q^2) + 2 G_m I^0(q^2) \label{mat21} \\
D_{22}(q^2) & = & 2 G_8 I^8(q^2) + 2 G_m I^m(q^2) - 1 \, . \label{mat22}
\end{eqnarray}
From the pole positions of the scattering amplitude Eq. (\ref{qas1}), the 
$\sigma$-meson mass $m_{\sigma}$ and the $f_0$-meson mass $m_{f_0}$ 
are determined.
\par
    The scattering amplitude Eq. (\ref{qas1}) can be diagonalized by rotation
in the flavor space 
\begin{eqnarray}
{\cal T}_{q \bar q} & = & -
\left(
\begin{array}{c} 
\bar u(p_3) \lambda^8 v(p_4) \\
\bar u(p_3) \lambda^0 v(p_4) 
\end{array} 
\right)^T   {\bf T}_{\theta}^{-1} {\bf T}_{\theta} 
\left(
\begin{array}{cc}
A(q^2) & B(q^2) \\
B(q^2) & C(q^2) \\
\end{array}
\right) {\bf T}^{-1}_{\theta}  \nonumber \\
&& \times {\bf T}_{\theta}  
\left(
\begin{array}{c}
\bar v(p_2) \lambda^8 u(p_1) \\
\bar v(p_2) \lambda^0 u(p_1)
\end{array}
\right) \, , \label{qasm1} \\
& = & -
\left(
\begin{array}{c} 
\bar u(p_3) \lambda^{\sigma} v(p_4) \\
\bar u(p_3) \lambda^{f_0}  v(p_4) 
\end{array} 
\right)^T \, 
\left(
\begin{array}{cc}
D^{\sigma}(q^2) & 0 \\
0 & D^{f_0}(q^2) 
\end{array}
\right) \nonumber \\
&& \times
\left(
\begin{array}{c}
\bar v(p_2) \lambda^{\sigma} u(p_1) \\
\bar v(p_2) \lambda^{f_0} u(p_1)
\end{array}
\right) \, , \label{qasm2}
\end{eqnarray}
with $\lambda^{\sigma} \equiv \cos \theta  \lambda^8 - \sin \theta \lambda^0$,
$\lambda^{f_0} \equiv \sin \theta  \lambda^8 + \cos \theta \lambda^0$ and
\begin{equation}
{\bf T}_{\theta} = \left(
\begin{array}{cc}
\cos \theta & -\sin \theta \\
\sin \theta & \cos \theta 
\end{array}
\right) \, .
\end{equation}
The rotation angle $\theta$ is determined by 
\begin{equation}
\tan 2 \theta = \frac{2 B(q^2)}{C(q^2) - A(q^2)} \, . \label{angle}
\end{equation}
Note that $\theta$ therefore depends on $q^2$.  At $q^2 = m_{\sigma}^2$,  
$\theta$ represents the 
mixing angle of the $\lambda^8$ and $\lambda^0$ components in the 
$\sigma$-meson state.

The \UA\ breaking KMT 6-quark determinat interaction ${\cal L}_6$ contributes
to the scalar $q \qbar$ channel only by the form of the effective 4-quark 
interaction, which is derived from ${\cal L}_6$ by contracting 
a quark-antiquark pair into the quark condensate. The explicit form of 
the effective KMT interaction is 
\bea
{\cal L}_6^{eff} =\left( \frac{-1}{2} \right) G_D \Biggl\{ & &
\left( \frac{-2}{3} \right) \left( 2 \la \ubar u \ra + \la \sbar s \ra \right)
\left[ \left(\bar \psi \lambda^0 \psi \right)^2 
     - \left(\bar \psi \lambda^0 i \gamma_5 \psi \right)^2 \right] 
\nonumber \\
& & + \la \sbar s \ra \sum_{i=1}^3 
\left[ \left(\bar \psi \lambda^i \psi \right)^2 
     - \left(\bar \psi \lambda^i i \gamma_5 \psi \right)^2 \right] 
\nonumber \\
& & + \la \ubar u \ra \sum_{i=4}^7 
\left[ \left(\bar \psi \lambda^i \psi \right)^2 
     - \left(\bar \psi \lambda^i i \gamma_5 \psi \right)^2 \right] 
\nonumber \\
& & + \left( \frac{1}{3} \right) 
\left( 4 \la \ubar u \ra - \la \sbar s \ra \right)
\left[ \left(\bar \psi \lambda^8 \psi \right)^2 
     - \left(\bar \psi \lambda^8 i \gamma_5  \psi \right)^2 \right] 
\nonumber \\
& & + \left( \frac{\sqrt{2}}{3} \right) 
\left( \la \ubar u \ra - \la \sbar s \ra \right) 
\Bigl[ \left(\bar \psi \lambda^0 \psi \right) 
       \left(\bar \psi \lambda^8 \psi  \right) 
+ \left(\bar \psi \lambda^8 \psi \right) 
       \left(\bar \psi \lambda^0 \psi  \right) 
\nonumber \\
& &  - \left(\bar \psi \lambda^0 i \gamma_5 \psi \right) 
       \left(\bar \psi \lambda^8 i \gamma_5 \psi  \right) 
+ \left(\bar \psi \lambda^8 i \gamma_5 \psi \right) 
       \left(\bar \psi \lambda^0 i \gamma_5 \psi  \right) \Bigr] 
\Biggr\} \, . \label{ekmt}
\eea
One can easily figure out from Eq. (\ref{ekmt}) that the \UA\ breaking KMT
interaction gives the attractive force in the flavor singlet scalar $q \qbar$ 
channel. On the other hand, it gives the repulsive force in the 
isospin $I = 1$ ($a_0$) and $I = 1/2$ ($K_0^*$) channels. 
Because of the large strange quark mass, $| \la \sbar s \ra |$ is bigger than 
$| \la \ubar u \ra |$, and therefore, the repulsion 
in the $I = 1$ channel is stronger than that in the $I = 1/2$ channel.
\section{Results}
We show our numerical results and give discussions on the mass 
spectrum as well as the mixings in this section.
As the extended NJL model has been used in the analyses of the pseudoscalar 
mesons, here we have used the model parameters fixed in the study of the 
electromagnetic decays of the $\eta$ meson. Since the $\eta$ meson 
properties depend on the strength of the \UA\
breaking interaction rather sensitively, 
it is reasonable to determine the strength of the
\UA\ breaking interaction from the $\eta$ meson properties.

The parameters of the NJL model are the current quark masses 
$m_{u}=m_{d}$, $m_{s}$, the four-quark coupling constant $G_{S}$, the 
\UA\ breaking KMT six-quark determinant coupling constant $G_{D}$ and the 
covariant cutoff $\Lambda$.  We take $G_{D}$ as a free parameter and study 
scalar meson properties as functions of $G_{D}$.
We use the light current quark masses $m_{u}=m_{d}=8.0$ MeV to reproduce 
$M_u = M_d \simeq 330$ MeV ($\simeq 1/3 M_N$) which is the value
commonly used in the constituent quark model.
The other parameters, $m_{s}$, $G_{S}$ and $\Lambda$, are determined
so as to reproduce the isospin averaged observed masses, $m_{\pi} = 138.0$ 
MeV, $m_{K} = 495.7$ MeV and the pion decay constant $f_{\pi} = 92.4$ MeV.
When we take the different value of $G_{D}$, we go through the fitting 
procedure each time.
\par
    We obtain $m_{s}=193$ MeV, $\Lambda=783$ MeV, $M_{u,d}=325$ MeV,
$M_{s}=529$ MeV and 
$f_K = 97$ MeV, which are almost independent of $G_{D}$.
The quark condensates are also independent of $G_D$ and our results are 
$\langle \bar uu \rangle^{\frac{1}{3}} = -216$ MeV and 
$\langle \bar ss \rangle^{\frac{1}{3}} = -226$ MeV whenever we have fixed 
other model parameters from the observed values of $m_\pi$, $m_K$ and $f_\pi$.
\par
    We define dimensionless parameters,
\bea
G_{D}^{\rm eff} &\equiv& - G_{D} (\Lambda / 2 \pi)^{4} \Lambda N_{c}^{2}
\nonumber\\
G_{S}^{\rm eff} &\equiv& G_{S} (\Lambda / 2 \pi)^{2} N_{c} .
\eea
As reported in Ref.~\cite{TNO1997}, the experimental value of the 
$\eta \to \gamma \gamma$ decay amplitude is reproduced at about 
$G_D^{\rm eff} = 0.7$. The calculated $\eta$-meson mass at 
$G_D^{\rm eff} = 0.7$ is $m_{\eta} = 510$ MeV which is 7\% smaller than 
the observed mass.  
$G_D^{\rm eff} = 0.7$ corresponds to 
$G_{D} \langle \overline{s} s \rangle / G_{S} =0.44$, suggesting that 
the contribution from ${\cal L}_{6}$ to the dynamical mass of 
the up and down quarks is 44\% of that from ${\cal L}_{4}$.
The calculated value of $\Gamma(\eta \to \pi^0 \gamma \gamma)$ is 0.92 eV 
at $G_D^{\rm eff} = 0.7$, which is in good agreement with the experimental 
data: $\Gamma(\eta \to \pi^0 \gamma \gamma) = 0.93 \pm 0.19$ eV.

Before going to present the numerical results for the scalar mesons, let us 
summarize the properties of the scalar mesons in the NJL model.
In the $SU_L(2) \times SU_R(2)$ version of the NJL with no explicit 
symmetry breaking term, the $\sigma$-meson mass can be calculated 
analytically in the mean field + ladder approximation, \ie, 
$m_\sigma = 2 M_u$.  
The $\sigma$ meson is therefore regarded as the lowest bosonic excitation, 
whose mass is twice of the gap energy, associated with chiral symmetry 
breaking.
It should be noticed that there is a cut above $q^2 = 4 M_u^2$ in the complex 
$q^2$-plane of the quark-antiquark scattering T-matrix, 
which corresponds to 
the unphysical decay: $\sigma \to \qbar q$. This is one of the known
shortcomings of the NJL model.  If one introduces a small symmetry breaking 
term, \ie, the current quark mass term, $m_\sigma$ moves up and gets the 
imaginary part corresponding to the $\sigma \to \qbar q$ decay 
\cite{TTKK1990}.
The pole position is in the 
second Riemann-sheet of the complex $q^2$-plane,  as is the case of ordinary
resonances. It means that the Argand diagram for the T-matrix makes a circular
resonance shape in the scalar $q \qbar$ channel.%
\footnote{The situation is quite different in the case of the vector meson. 
In the nonrelativistic limit, the scalar meson channel corresponds to the 
p-wave quark-antiquark state whereas the vector meson channel corresponds to 
the s-wave quark-antiquark state. See Ref.~\cite{TKM1991}.}

It should be noted that the physical decay mode of $\sigma$, \ie,
$\sigma\to\pi\pi$ is neither taken into account in the ladder approximation.
As this decay makes the $\sigma$ width significantly large, our result for 
$\sigma$ mass is qualitative rather than quantitative.
Nevertheless, the results shown below show that the scalar mesons in 
the NJL model is realized as the chiral partner of the 
Nambu-Goldstone bosons, and that they give systematic behavior for 
the orders of the masses and the splittings.

\begin{figure}[t]
	\centerline{\includegraphics[width=10 cm]{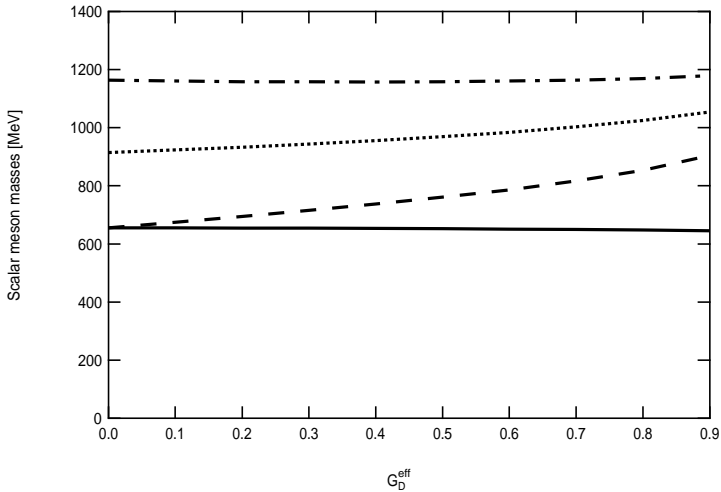}}
\caption{The calculated scalar meson masses as functions of the effective
coupling constant $G_D^{\rm eff}$ of the \UA\ breaking KMT interaction.
The solid, dashed, dotted and dash-dotted lines represent 
$m_\sigma$, $m_{a_0}$, $m_{K_0^*}$ and $m_{f_0}$, respectively.}
\label{fig:mass}
\end{figure}

\begin{figure}[t]
	\centerline{\includegraphics[width=10 cm]{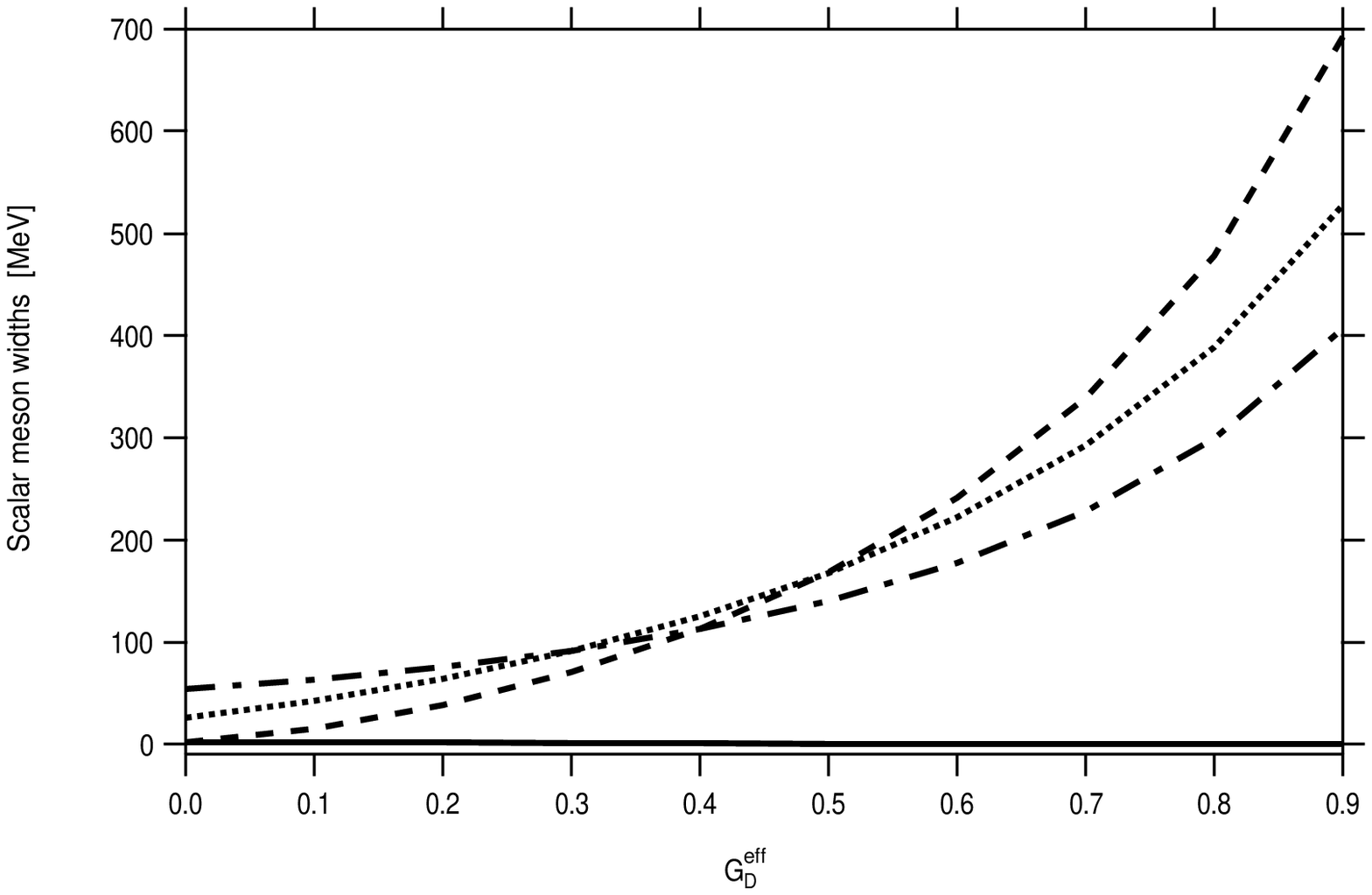}}
\caption{The calculated $q\qbar$ decay widths of the scalar mesons 
as functions of the effective coupling constant $G_D^{\rm eff}$ of 
the \UA\ breaking KMT interaction.
The solid, dashed, dotted and dash-dotted lines represent 
$m_\sigma$, $m_{a_0}$, $m_{K_0^*}$ and $m_{f_0}$, respectively.}
\label{fig:width}
\end{figure}

\begin{figure}[t]
	\centerline{\includegraphics[width=10 cm]{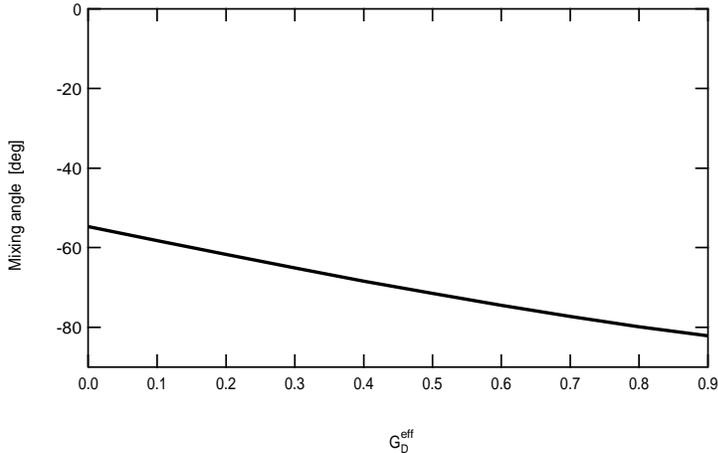}}
\caption{The calculated mixing angle of the $\sigma$ meson as 
a function of the effective coupling constant $G_D^{\rm eff}$ of 
the \UA\ breaking KMT interaction.}
\label{fig:mix}
\end{figure}

Let us now discuss our results of the scalar mesons. 
Here we call the lowest scalar meson states in the $I = 0, \, 1, \, 1/2$ 
channels and the second lowest one in the $I = 0$ channel 
$\sigma$, $a_0$, $K_0^*$ and $f_0$, respectively.  
The identification of these states with the 
experimentally observed states will be given in the next section.
The calculated results of the scalar-meson masses, $q\qbar$ decay widths and 
the mixing angle $\theta$ are shown in Fig. \ref{fig:mass}, 
Fig. \ref{fig:width} and Fig. \ref{fig:mix}, respectively.
The $q\qbar$ decay widths of the scalar mesons shown there are 
unphysical ones.  We present them just for showing the 
pole positions in the complex $q^2$-plane.
When $G_{D}^{\rm eff}$ is zero, our lagrangian does not cause the flavor 
mixing and therefore the ideal mixing is achieved. 
The $\sigma$ is purely $u\bar u + d\bar d$, which corresponds to 
$\theta = -54.7^{\circ}$, and is degenerate to the $a_0$ in this limit.
When one increases the strength of the \UA\ breaking KMT interaction, 
the $q\qbar$ attraction in $\sigma$ increases and $\sigma$ state moves 
from the ideal mixing state toward the flavor singlet state. 
It means that the strange quark component of $\sigma$ increases as 
$G_D^{\rm eff}$ becomes larger.  
Since the increase of the attractive force compensates with the increase 
of the strange quark component, 
$m_\sigma$ is almost independent of the strength of the \UA\ breaking 
interaction and our result is $m_\sigma =650$ MeV at $G_{D}^{\rm eff} = 0.7$. 
The $\qbar q$ decay width of the $\sigma$ meson is very small, i.e., 
less than 2 MeV and therefore we neglect it in our calculation of 
the mixing angle.
At $G_{D}^{\rm eff} = 0.7$, the calculated mixing angle is
$\theta = -77.3^\circ$, corresponding to  about 15\% mixing of 
the strangeness component in $\sigma$.

Hatsuda and Kunihiro have discussed the masses and mixing angle of the 
isoscalar nonstrange ($\sigma_{NS}$) and strange ($\sigma_S$) scalar mesons 
using the similar model \cite{HK1994}. 
They have reported a rather small mixing 
between $\sigma_{NS}$ and $\sigma_S$. The reason of the difference between 
their result and our result is the strength of the $U_A(1)$ breaking KMT
interaction.  The strength of the $U_A(1)$ breaking KMT interaction used in 
the present study is much stronger than that used in their study. 
They have determined the strength from the $\eta'$ mass, while we have 
fixed it from the radiative decays of $\eta$. Strong $U_A(1)$ breaking
interaction suggests that the instanton liquid picture of the QCD vacuum
\cite{Shuryak1982}.
In Ref.~\cite{HK1994}, they have discussed the origin of the difference of
the mixing properties of the scalar mesons and pseudoscalar mesons. 
We agree with their qualitative discussion, namely, the flavor mixing 
between $\sigma$ and $f_0$ is weaker than that between 
$\eta$ and $\eta'$.
Shakin has also pointed out that the KMT interaction mixes the 
$\sigma_{NS}$ and $\sigma_S$, while he assigned the lowest $I=0$ $q\qbar$ 
state to $f_0(980)$ \cite{Sha2002}.

Let us turn to the discussion of the $a_0$ and $K_0^*$ mesons.
As shown in  Fig. \ref{fig:mass}, both $m_{a_0}$ and $m_{K_0^*}$ 
increase as $G_D^{\rm eff}$ increases. The slope for $m_{a_0}$ is 
steeper than that for $m_{K_0^*}$, 
which is consistent with the simple argument based 
on the form of the effective interaction Eq. (\ref{ekmt}).
At $G_{D}^{\rm eff} = 0.7$, the calculated masses are $m_{a_0} = 816$ MeV and
$m_{K_0^*} = 1002$ MeV, therefore
the \UA\ breaking interaction pushes up the $a_0$ and $K_0^*$ masses about 
161 MeV and 88 MeV, respectively. 
Although the effect of the \UA\ breaking interaction on the $K_0^*$ meson 
is smaller than that on the $a_0$ meson, our numerical results show that 
it is not enough to support the existence of the light $\kappa$ state.

As for the $f_0$ meson, we have shown our results in 
Figs. \ref{fig:mass} and \ref{fig:width}.  At $G_{D}^{\rm eff} = 0$, 
the $f_0$ state is expected to be pure $s\sbar$ state in our model.
Because of the $q\qbar$ decay width, we cannot calculate the 
mixing angle for $f_0$.  The calculated mass of the $f_0$ meson 
at $G_{D}^{\rm eff} = 0$ is $m_{f_0} = 1.163$ GeV which is above the 
$s\sbar$ threshold $2 M_s = 1.113$ GeV. As shown in Ref.~\cite{Dm1996}, 
the symmetry breaking effect by the current quark mass term pushes up the 
scalar meson mass above the $q\qbar$ threshold and the following relation 
is obtained by using the bosonization technique with the lowest order
derivative expansion in the NJL model.
\begin{equation}
\left( m_{\rm scalar \, meson}^2 - ({q\bar{q} \, {\rm threshold \,energy})}^2 
\right) 
\propto m_{\rm current \, quark}
\end{equation}
Our results at $G_{D}^{\rm eff} = 0$ are $m_{a_0}^2 - 4 M_u^2 = 0.008$ 
GeV$^2$, 
$m_{K_0^*}^2 - (M_u + M_s)^2 = 0.060$ GeV$^2$ and 
$m_{f_0}^2 - 4 M_s^2 = 0.115$ GeV$^2$, respectively.  The above 
simple mass relation therefore holds in our case too.
Fig. \ref{fig:mass} shows that the $f_0$ meson mass is almost independent
of the strength of the KMT interaction. The situation is just opposite to the 
$\sigma$ case, i.e., the increase of the repulsive force by the KMT interaction
compensates with the decrease of the strange quark component of the $f_0$ meson
when one increases the strength of the KMT interaction.

It should be noted here that 
in the $SU_L(3) \times SU_R(3)$ version of 
linear sigma model, not only the three-meson flavor 
determinant term but also the chiral invariant 
four-meson terms give rise to the $\sigma - a_0$ mass difference
\cite{CH1974,BFMNS2001}.
We note that the extended NJL model does not give such type of interaction.

%===========================================================
\section{Conclusion}
%===========================================================
The aim of this study is to show the roles of two important symmetry
structures of QCD, \ie, the \UA\ anomaly
as well as the dynamical chiral symmetry breaking in the spectrum of 
the light scalar mesons.  We suggest that the anomalous ordering of the
$\sigma-a_0$ is understood by these features.
To this end, we have studied the effects of the \UA\ breaking 
interaction on the low-lying 
nonet scalar mesons using the extended Nambu-Jona-Lasinio model. 
The strength of the \UA\ breaking interaction has been determined by the 
electromagnetic decays of the $\eta$ meson and we have obtained 
rather strong \UA\ breaking interaction, which reminds us of the 
instanton liquid picture of the QCD vacuum. 

We have found that the 
\UA\ breaking interaction gives rise to about 150 MeV
mass difference between the $\sigma$ and $a_0$ mesons. 
We obtain the low-lying $I=0$ scalar meson as the chiral partner of the 
pion with mass about 650 MeV. We identify it with 
$\sigma(600)$. In our present scheme, 
it is the $q\qbar$ state.
We have also found that the strangeness content in the $\sigma$ meson is 
about 15\%.  The flavor contents of the $\sigma$ meson may be observed in 
the analysis of the $J/\psi$ decays.  
Furthermore, since the $\sigma$ meson plays the central role in the 
intermediate range attraction of the nuclear force, the strange quark 
content in the $\sigma$ meson may be important in the context of 
the hyperon-hyperon interactions.

As for the $I=1$ channel, we find the $q\qbar$ resonance state at 
$m = 816$ MeV.  We identify it with $a_0(980)$ though the mass is 
still smaller than the observed value.
In the $I=1/2$ channel, we find  the $q\qbar$ resonance state at 
$m = 1002$ MeV, which is above the $I=1$ state. 
A possible reason that the $a_0$ mass is not large enough is the large
unphysical $q\qbar$ decay width shown in Fig. \ref{fig:width}, 
which is an artifact of this model.
On the other hand, the order of the $a_0 - K_0^*$ masses is not likely to
change within this model.  Because the unphysical $q\qbar$ widths are 
similar for $a_0$ and $K_0^*$, this conclusion may not be affected by 
this artifact.  We therefore consider that the obtained $K_0^*$ is not 
identified with recently reported $\kappa(700-900)$.
A possible candidate is $K_0^*(1430)$, although the mass difference
between $K_0^*(1430)$ and $a_0(980)$ is rather large. It seems to be 
difficult to explain it from the symmetry breaking effect by the 
current strange quark mass.
A possible scenario to explain this discrepancy is that the coupling of 
the $K \pi$ channel happens to be so large that its mixing leads to two scalar 
states $\kappa(700-900)$ and $K_0^*(1430)$.
As for the second lowest state in the $I=0$ channel, the pole appears 
at $m = 1164$ MeV, which is above the $s\sbar$ threshold and is about 
350 MeV heavier than the mass of our $I=1$ state. 
We therefore consider it may not be the $f_0(980)$ state. 
The possible candidates are $f_0(1370)$ and $f_0(1500)$.
They again may be the results of mixings of two meson states, 
such as $K \bar K$, as well as glueball states.

We should note that the NJL model is a crude effective model of 
QCD with some shortcomings.  Especially, as the model does not provide
confinement of quarks, the scalar mesons are not free from $q \qbar$ decay 
channel.  
Therefore, the numerical results obtained in this paper should be taken 
at most qualitatively.
Yet, we demonstrate that the significant $\sigma-a_0$ mass difference is 
induced by the \UA\ anomaly.  Guided by this success, it is most desirable 
to confirm this mechanism directly in QCD.  
For instance, the lattice QCD calculation with controlled topological charge 
density and/or the QCD sum rule with direct instanton effects are 
future possibilities.
Also a study of the nonet scalar mesons using the more realistic quark model 
approaches, such as the improved ladder model of QCD\cite{NNTYO2000}, 
will give us further confidence on the mechanism and structure of 
the scalar meson spectrum.
Such work is underway and is to be published elsewhere {\cite{Umekawa_D}}.

\section*{Acknowledgements}
We are happy to thank Drs. Teiji Kunihiro and Muneyuki Ishida for helpful 
discussions.
This work is supported in part by the Grant-in-aid for Scientific Research
(C) (2) 11640261 of the Ministry of Education, Science, Sports and Culture
of Japan.

\end{document}